\documentclass[fleqn,usenatbib]{mnras}
\usepackage{orcidlink}

\usepackage{url,hyperref,microtype,subcaption}

\usepackage{orcidlink}
\usepackage{verbatim}

 \usepackage{amsmath}
\usepackage{xcolor}

\definecolor{my_color}{HTML}{3a18b1}

\definecolor{new_color}{HTML}{CF0000}
\definecolor{new_black}{HTML}{000000}
\definecolor{mycitecolor}{HTML}{b752ba}
\definecolor{mycitecolor2}{HTML}{8400c2}
\newcommand\bedit{\textcolor{new_black}}

\newcommand{\Spitzer}{{\it Spitzer}}
\newcommand{\TESS}{{\it TESS}}

\newcommand{\be}{\begin{equation}}
\newcommand{\ee}{\end{equation}}

\newcommand{\msun}{M\ensuremath{_\odot}}

\newcommand{\mj}{M$_J$}

\usepackage{newtxtext,newtxmath}

\usepackage[T1]{fontenc}

\DeclareRobustCommand{\VAN}[3]{#2}
\let\VANthebibliography\thebibliography
\def\thebibliography{\DeclareRobustCommand{\VAN}[3]{##3}\VANthebibliography}

\usepackage{graphicx}	
\usepackage{amsmath}	

\title[How Lonely is WD 1856 b?]{TTV Constraints on Additional Planets in the WD 1856+534 system}

\author[Kubiak et al.]{
Sarah Kubiak\,$^{1,2,*}$ \orcidlink{0000-0001-6699-742X}, Andrew Vanderburg\,$^{3}$ \orcidlink{0000-0001-7246-5438}, Juliette Becker\,$^{4,2,5}$ \orcidlink{0000-0002-7733-4522}, Bruce Gary\,$^{6}$\orcidlink{0000-0002-4080-1342}, Saul A. Rappaport\,$^{3}$\orcidlink{0000-0003-3182-5569},\newauthor Siyi Xu\,$^{7}$\orcidlink{0000-0002-8808-4282}, Zoe de Beurs\,$^{8,9}$\orcidlink{0000-0002-7564-6047}
\\
$^{1}$Department of Journalism and Media Communication, Colorado State University, Fort Collins, CO, 80521, USA \\
$^{2}$Department of Astronomy, University of Wisconsin-Madison, 475 N. Charter St., Madison, WI 53703, USA \\
$^{3}$Department of Physics and Kavli Institute for Astrophysics and Space Research, Massachusetts Institute of Technology, Cambridge, MA 02139, USA\\
$^4$ Division of Geological and Planetary Sciences, California Institute of Technology, Pasadena, CA 91125, USA\\
$^5$ 51 Pegasi b Fellow\\
$^6$ Hereford Arizona Observatory, Hereford, AZ 85615, USA \\
$^{7}$ Gemini Observatory/NSF’s NOIRLab, 670 N. A’ohoku Place, Hilo, Hawaii, 96720, USA\\
$^8$ Department of Earth, Atmospheric and Planetary Sciences, Massachusetts Institute of Technology, Cambridge, MA 02139, USA\\
$^9$ NSF Graduate Research Fellow and MIT Presidential Fellow
}

\date{Accepted XXX. Received YYY; in original form ZZZ}

\pubyear{2022}

\begin{document}
\label{firstpage}
\pagerange{\pageref{firstpage}--\pageref{lastpage}}
\maketitle

\begin{abstract}
WD 1856+534 b (or WD 1856 b for short) is the first known transiting planet candidate around a white dwarf star. WD 1856 b is about the size of Jupiter, has a mass less than about 12 Jupiter masses, and orbits at a distance of about 2\% of an astronomical unit. The formation and migration history of this object is still a mystery. Here, we present constraints on the presence of long-period companions (where we explored eccentricity, inclination, mass, and period for the possible companion) in the WD 1856+534 planetary system from Transit Timing Variations (TTVs). We show that existing transit observations can rule out planets with orbital periods less than about 500 days. With additional transit observations over the next decade, it will be possible to test whether WD 1856 also hosts additional long-period planets that could have perturbed WD 1856 b into its current close-in orbit. 
\end{abstract}




\section{Introduction}

The past three decades have ushered in an exoplanet revolution with the discovery of over 5000 exoplanets \citep{akeson}. Astronomers have learned that exoplanets are common in our galaxy \citep[e.g.][]{kunimoto} and show a greater diversity in sizes \citep{barclay, weiss2013, zhou}, compositions \citep{masuda, santerne}, and architectures \citep{lissauer, becker, bourrier} than the planets in our own solar system. However, almost all of the exoplanets known today orbit main sequence stars that are still actively burning hydrogen into helium. Eventually, each of these main-sequence planet-hosting stars will \bedit{undergo} a dramatic transformation at the end of their lifetimes that will have major impacts on their orbiting planetary systems. 

A white dwarf is the remnant core of a low-mass ($M_\star \lesssim 8 \msun$) star at the end of its life, after it runs out of hydrogen fuel. When a star transitions from burning hydrogen on the main sequence into an inert white dwarf, it first puffs up into a red giant, more than a hundred times its initial size.  As the red giant expands, it is expected to engulf and evaporate  planetary material \citep[except for the most massive objects][]{passy} and carve out a zone without surviving planets in the inner region of the solar system \citep{nordhaus}.  It will also likely destroy many small asteroids within a few AU of the host star \citep{Jura2008}.   


As a result of the red giant phase of evolution, exoplanets orbiting close to white dwarf stars are expected to be rare, which is why the discovery of a close planetary-mass companion to the white dwarf WD 1856+534 was surprising \citep{Vanderburg2020}. \bedit{WD 1856} is a white dwarf star that was found to host a transiting companion in data from the \textit{Transiting Exoplanet Survey Satellite} (\textit{TESS}) mission \citep{ricker}. \citet{Vanderburg2020} detected periodic dips in the \TESS\ light curve of WD 1856 indicating a close-in companion orbiting the star every 1.4 days called WD 1856 b. Follow-up observations from ground-based telescopes and the \Spitzer\ space telescope showed that the companion has a size similar to Jupiter, and a mass less than about 12 Jupiter masses.  This planetary system is intriguing because the closely orbiting Jupiter-sized planet could not have existed in its current location while the host star was on the main sequence, or it would have been engulfed. Instead, the planet must have orbited farther away from the star, and migrated into its current location either during or after the star's evolution into a white dwarf. 


The evolution of WD 1856 b's orbit is still uncertain and the subject of significant debate. There are two main mechanisms that could plausibly lead to the planet's current orbit: 1) WD 1856 b either could have survived engulfment by the red giant \bedit{and} emerge at its current orbit intact \citep{lagos, chamandy, merlov}, or 2) it could have migrated via tidal circularization after being driven to a high eccentricity orbit by other objects in the system (either other planets \citealt{maldonado, oconnor2} or a pair of distant M-dwarf companions \citealt{munoz, oconnorO, stephan}). 

The first option is called ``common envelope evolution.'' In this scenario, the expanding star grows large enough to engulf a lower mass binary companion, and friction causes it to rapidly spiral inwards while depositing orbital energy into the giant star's envelope. If the companion is able to deposit enough gravitational potential energy, then the envelope is ejected and the companion's inward migration is halted. If there is insufficient gravitational potential energy to unbind the envelope then the companion and the stars merge. WD 1856 b has a low mass and long orbital period compared to other post common envelope binaries \citep{Vanderburg2020} which makes it challenging to explain its formation by this mechanism, but it could be possible with additional energy sources like recombination energy in the envelope or energy deposited by other planets in the system \citep{lagos, chamandy, merlov}. 

The second scenario that could produce WD 1856 b's current orbit is high eccentricity migration, one of the main ways by which hot Jupiters around main sequence stars are believed to have arrived in their current orbits \citep{dawsonjohnson}. In this scenario, gravitational interactions between WD 1856 b and other bodies in the system increase WD 1856 b's orbital eccentricity. At the planet's closest approach to the star it is distorted by tidal forces which dissipates a small amount of orbital energy and decreases the orbital semimajor axis. Over the course of $\sim10^8-10^9$ years of tidal dissipation, the planet could slowly migrate into its present-day close-in orbit. One way the planet could reach such high eccentricities is via the von Zeipel-Kozai-Lidov (ZKL) mechanism \citep{vonzeipel,kozai,lidov}, a secular dynamical process that causes oscillations in orbital inclination and eccentricity in the presence of distant, orbitally misaligned companions. WD 1856 is orbited by two M-dwarfs at a distance of $\approx$1500 AU \citep{Vanderburg2020} that could trigger ZKL oscillations \citep{munoz, oconnorO, stephan}.  Another way WD 1856 b could reach high eccentricity is through interactions with other planets in the system, if they exist. In the presence of additional Jupiter-mass planets, WD 1856 b could plausibly reach high enough eccentricities to tidally circularize, but it is  unknown if WD 1856 b has had enough time to do this \citep{Vanderburg2020, maldonado, oconnor2}. 


Currently, astronomers are planning and performing follow-up observations of the WD 1856 system to better understand the system's formation. The main goal of most of the follow-up observations planed so far is to detect spectroscopic features in the planet's atmosphere, and use these features to constrain the planet's mass. A precise measurement of the planet's mass would help differentiate among the different formation scenarios, since common envelope evolution likely requires more massive companions than high-eccentricity migration \citep{Xu2021}. Recently, \citet{RAlonso2021} and \citet{Xu2021} reported transmission spectroscopy observations to constrain the planet's mass. These observations, carried out on 10-m class ground-based telescopes, detected no statistically significant features in the planet's transmission spectrum, and yielded a lower limit on the mass of about 0.84 Jupiter mass \citep{Xu2021}. Meanwhile, two programs with the \textit{James Webb Space Telescope} have been approved for Cycle 1 observations \citep{jwst1,jwst2} to help constrain the planet's mass and study its atmosphere. 

In this work, we take a different approach to try to shed light on WD 1856 b's present orbit: searching for additional planets in the system. \bedit{If the system reached its current orbit} via high eccentricity migration excited by gravitational interactions with other planets, there should be other gas giant planets still orbiting in the system (unless they were ejected in the ensuing dynamical chaos). We search for planets by measuring the times of many transits of WD 1856 b to see whether the planet sometimes transits slightly earlier or later than expected. Such changes are called transit timing variations\footnote{\bedit{Traditionally in exoplanet literature, TTVs refer to actual changes in planets' orbits due to dynamical perturbations that cause the transits to happen earlier or later. Here we also include effects such as the light travel time effect or Roemer delay that are usually ignored in mainstream TTV analysis. }} (TTVs) and have been used to detect previously unknown planets \citep[e.g.][]{Ballard2011} and measure the masses of known planets \citep{Carter2012}. TTV measurements have also been used to detect planet candidates around eclipsing white dwarf/M-dwarf binary stars \citep[e.g.][]{Marsh2014}, but it is challenging to interpret these variations because magnetic fields in the M-dwarf companions \citep{applegate} often seem to drive changes in the binary orbit that can mimic orbital variations caused by additional planets \bedit{\citep{bours2016, Pulley2022}}. Since WD 1856 b is much lower in mass than the M-dwarf companions that often show orbital variations, it may be possible to circumvent this problem, although any detected signals must still be subjected to the highest scrutiny for these reasons.

Our paper is organized as follows. In Section \ref{observations}, we describe how we obtained the transit timing data for WD 1856 b. In Section \ref{Analysis} we discuss our process for modeling the TTVs and placing limits on the presence of planets. In Section \ref{Results} we summarize the results of our analysis, and in Section \ref{discussion} we discuss its implications and conclude.

\begin{figure*}
\centering
\includegraphics[width=1.0\textwidth]{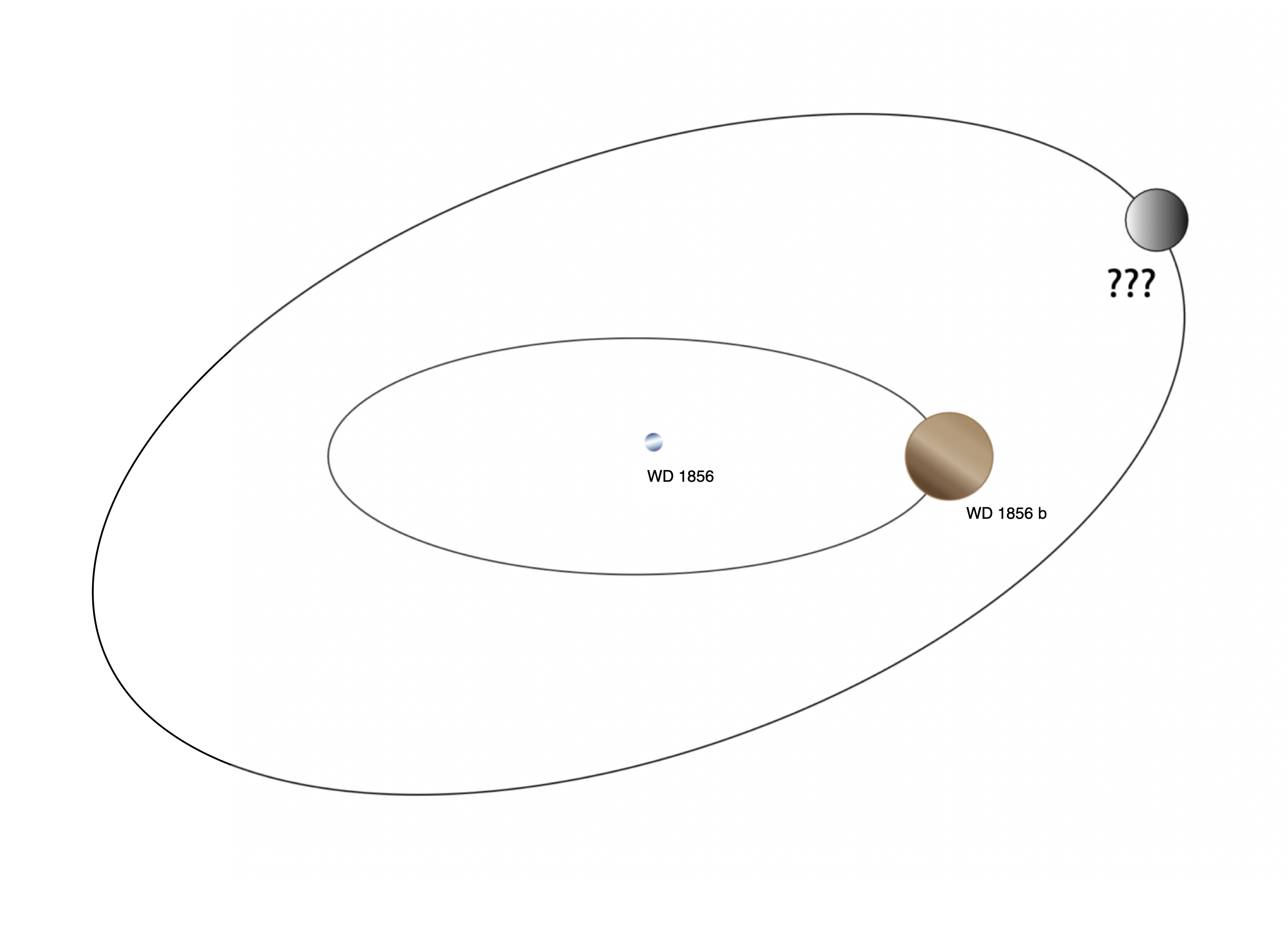}
\caption{A schematic diagram of the WD 1856 system. Shown here are the white dwarf WD 1856, the transiting companion WD 1856 b, and a hypothetical outer companion planet -- the subject of our study. In this work, we use transit timing variations to search for, and rule out, plausible properties of hypothetical companion planets and explore how these constraints depend on the hypothetical companion's mass, orbital period, eccentricity, and inclination. Ultimately, constraints such as these will be useful to assessing whether planets exist in the WD 1856 system that could have caused the inner planet's migration by planet/planet scattering. }\label{schematic}
\end{figure*}

\section{Observations}\label{observations}

\begin{figure*}
\centering
\includegraphics[width=\textwidth]{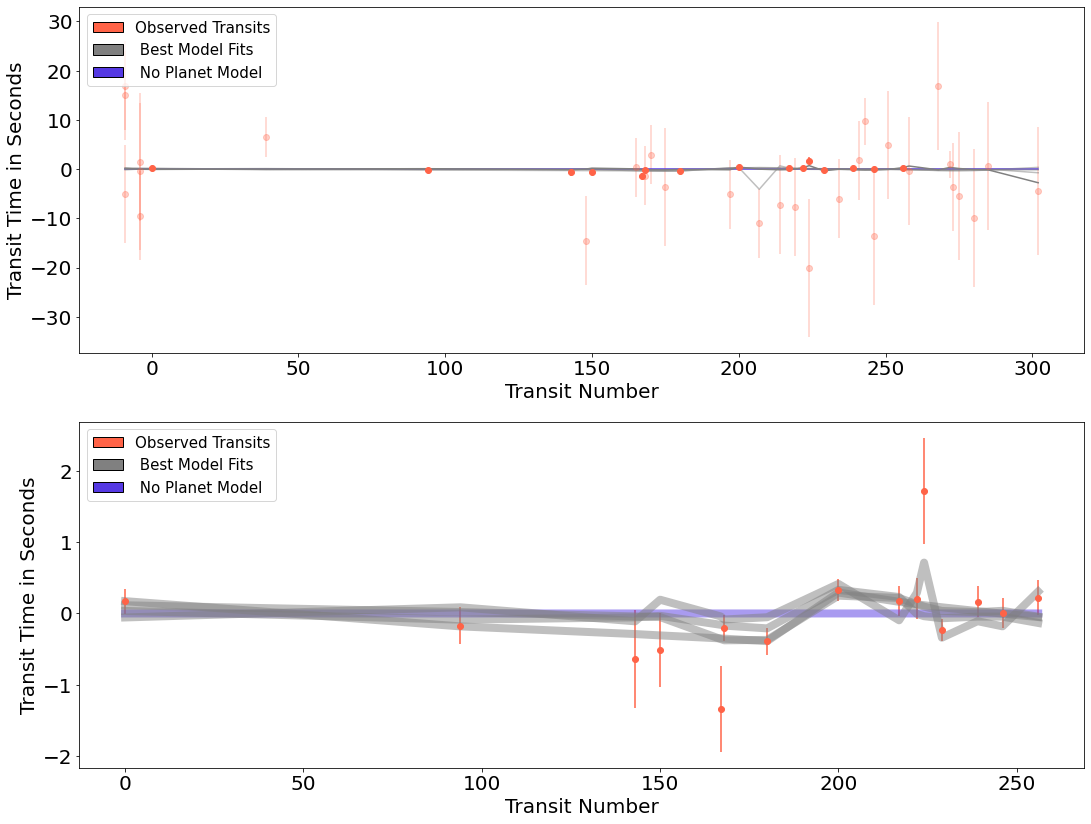}
\caption{ Measured transit times for WD 1856 b, along with a handful of best-fit models from our various analyses described in Sections \ref{Analysis} and \ref{Results}. The data points shown are the residuals of a linear fit to the measured transit times, and therefore show any deviations from perfect periodicity. The dark points represent highly precise measurements from either the GTC or Gemini telescopes, and fainter points represent the rest of the observations listed in Table \ref{transittimes}. The nominal model (with no timing variations) is shown as a blue line, while an assortment of best-fit two-planet models from various slices through parameter space (see Section \ref{planetlimits}) are shown in grey. The top panel shows all of the transits used in our analysis, while the bottom panel shows a zoomed-in view of the most precisely measured transit times.  }\label{ttvs}
\end{figure*}

For our analysis in this work, we use a combination of previously published transit times and newly measured transit times. The transit times we used in this paper are listed in Table \ref{transittimes} and their deviations from a perfectly periodic model are shown in Figure \ref{ttvs}.  In this section, we document the sources of the transit times we use in our analysis. 

\begin{table}
	\begin{center}
	\caption{Transit Timing Observations of WD 1856 b}
\label{transittimes}
\scriptsize
	\begin{tabular}{llll} 
		\hline
		Transit & Time (\bedit{BJD\_TBD}) and & Observatory & Reference\\
		Number & Uncertainty (days) & & \\
		\hline
-4 & 2458773.74321663 $\pm$ 1.0e-04 & HAO & \citet{Vanderburg2020}\\ 
-4 & 2458773.7433208 $\pm$ 1.9e-04 & JBO & \citet{Vanderburg2020}\\ 
-4 & 2458773.74334395 $\pm$ 1.4e-04 & RVO & \citet{Vanderburg2020}\\ 
-9 & 2458766.703572 $\pm$ 1.2e-04 & HAO & \citet{Vanderburg2020} \\ 
-9 & 2458766.70380349 $\pm$ 1.0e-04 & JBO & \citet{Vanderburg2020}\\ 
-9 & 2458766.70382663 $\pm$ 1.0e-04 & RVO & \citet{Vanderburg2020} \\ 
0 & 2458779.375085 $\pm$ 2.0e-06 & GTC & \citet{Vanderburg2020}\\ 
39 & 2458834.284788 $\pm$ 4.7e-05 & \Spitzer\ & \citet{Vanderburg2020} \\ 
94 & 2458911.721367 $\pm$ 3.0e-06 & GTC & \citet{RAlonso2021}\\ 
143 & 2458980.710383 $\pm$ 8.0e-06 & GTC & \citet{RAlonso2021}\\ 
148 & 2458987.749918 $\pm$ 1.0e-04 & HAO & this work\\ 
150 & 2458990.565959 $\pm$ 6.0e-06 & GTC & \citet{RAlonso2021}\\ 
165 & 2459011.68505697 $\pm$ 6.9e-05 & HAO & this work\\ 
167 & 2459014.500916 $\pm$ 7.0e-06 & GTC & \citet{RAlonso2021}\\
168 & 2459015.90885532 $\pm$ 6.9e-05 & HAO & this work\\ 
168 & 2459015.90886837 $\pm$ 2.1e-06 & Gemini & \citet{Xu2021} \\ 
170 & 2459018.72478262 $\pm$ 6.9e-05 & HAO & this work\\ 
175 & 2459025.7644041 $\pm$ 1.4e-04 & HAO & this work\\ 
180 & 2459032.80413679 $\pm$ 2.2e-06 & Gemini & \citet{Xu2021}\\ 
197 & 2459056.7390488 $\pm$ 8.1e-05 & HAO & this work\\ 
200 & 2459060.96292936 $\pm$ 1.8e-06 & Gemini & \citet{Xu2021}\\ 
207 & 2459070.81837278 $\pm$ 8.1e-05 & HAO & this work\\ 
214 & 2459080.673991 $\pm$ 1.2e-04 & HAO & this work\\ 
217 & 2459084.89789415 $\pm$ 2.4e-06 & Gemini & \citet{Xu2021}\\ 
219 & 2459087.71368193 $\pm$ 1.2e-04 & HAO & this work\\ 
222 & 2459091.93759056 $\pm$ 3.3e-06 & Gemini & \citet{Xu2021}\\ 
224 & 2459094.75323397 $\pm$ 1.6e-04 & HAO & this work\\ 
224 & 2459094.7534865 $\pm$ 8.7e-06 & Gemini & \citet{Xu2021}\\ 
229 & 2459101.79315997 $\pm$ 1.8e-06 & Gemini & \citet{Xu2021}\\ 
234 & 2459108.83278943 $\pm$ 9.3e-05 & HAO & this work\\ 
239 & 2459115.87255671 $\pm$ 2.5e-06 & Gemini & \citet{Xu2021}\\ 
241 & 2459118.68845395 $\pm$ 9.3e-05 & HAO & this work\\ 
243 & 2459121.504424 $\pm$ 5.5e-05 & STELLA & \citet{stella} \\ 
246 & 2459125.72797126 $\pm$ 1.6e-04 & HAO & this work\\ 
246 & 2459125.72812932 $\pm$ 2.5e-06 & Gemini & \citet{Xu2021}\\ 
251 & 2459132.7678821 $\pm$ 1.3e-04 & HAO & this work\\ 
256 & 2459139.80752397 $\pm$ 2.8e-06 & Gemini & \citet{Xu2021}\\
258 & 2459142.62339615 $\pm$ 1.3e-04 & HAO & this work\\ 
268 & 2459156.70298634 $\pm$ 1.5e-04 & HAO & this work\\ 
272 & 2459162.33456 $\pm$ 3.1e-05 & STELLA & \citet{stella} \\ 
273 & 2459163.74244578 $\pm$ 1.0e-04 & HAO & this work\\ 
275 & 2459166.55830363 $\pm$ 1.5e-04 & HAO & this work\\ 
280 & 2459173.59794826 $\pm$ 1.6e-04 & HAO & this work\\ 
285 & 2459180.6377665 $\pm$ 1.5e-04 & HAO & this work\\ 
302 & 2459204.57267399 $\pm$ 1.5e-04 & HAO & this work\\

		\hline
	\end{tabular}
	\end{center}
\end{table}

\subsection{Literature Observations}\label{lightcurve}

Since its discovery, WD 1856 b has been observed by several different groups. We collected observations from the literature from three different papers in particular. 

First, we used transit observations collected by \citet{Vanderburg2020}. They observed WD 1856 b with a number of different telescopes, including \TESS, the Gran Telescopio Canarias (GTC) on the island of La Palma, Spain, the Telescopio Carlos S\'anchez (TCS) on the island of Tenerife, Spain, the Spitzer Space Telescope, and three privately owned telescopes in Arizona, USA: one at Raemor Vista Observatory (RVO), one at Junk Bond Observatory (JBO), and one at Hereford Arizona Observatory (HAO). We used all of these observations except for the \TESS\ data, which have much lower signal-to-noise per transit than the other observations, and the TCS observation, which was conducted simultaneously with the GTC observation, but with significantly lower signal-to-noise. \citet{Vanderburg2020} did not report individual transit times for their observations, so we used the transit times derived by \citet{RAlonso2021} for the GTC and Spitzer data, and we report transit times for the Raemor Vista, Junk Bond, and Hereford observations following the procedure described in Section \ref{haodata}. 

We also used transit times measured and reported by \citet{RAlonso2021}. They observed transits of WD 1856 b on several occasions with the GTC and derived transit times. Their observations were conducted using two different instruments and in different photometric bandpasses to support their primary science objective of measuring WD 1856 b's transmission spectrum. In particular, \citet{RAlonso2021} collected one photometric observation with the GTC’s OSIRIS instrument on March 2, 2020, one infrared observation with the GTC’s EMIR instrument on June 13 2020, and three observations with OSIRIS in a spectroscopic mode on May 3, 2020, May 10, 2020, and May 20, 2020. \citet{RAlonso2021} also derived transit times from archival transit observations collected by \citet{Vanderburg2020} from the GTC and \Spitzer. In total, they presented 7 precisely measured transit times, all of which have uncertainties smaller than 6 seconds, and 5 of which have uncertainties smaller than 1 second.

We used two observations of WD 1856 b collected by \citet{stella} using the 1.2 meter STELLar Activity (STELLA) telescope at on the island of Tenerife, Spain. These observations were conducted using the Wide Field STELLA Imaging Photometer (WiFSIP) imager on the 1.2 meter STELLA in Sloan i-band. We use the transit times measured by \citet{stella} in our analysis, which have uncertainties of 3-5 seconds. 

Finally, we used transit times measured by \citet{Xu2021} from the Gemini North telescope on Mauna Kea, Hawaii, USA. \citet{Xu2021} observed 10 transits of WD 1856 b using the GMOS multi-object spectrograph, also with the primary science goal of measuring the planet's transmission spectrum. These observations were taken on June 15, July 02, July 30, August 23, August 30, September 02, September 09, September 23, October 03, and October 17, all in 2020. \citet{Xu2021} derived transit parameters, including times, for each of the individual observations. For their transmission spectrum, \citet{Xu2021} excluded data from two transits that were taken in worse weather conditions, but we found that the transit times derived from these observations were still useful, albeit with somewhat larger uncertainties than the rest of the observations. All 10 Gemini observations have transit time uncertainties less than 1 second, and all but the two excluded by \citet{Xu2021} have uncertainties smaller than 0.25 second. 




\subsection{New Observations from Hereford Arizona Observatory}\label{haodata}

We obtained our own observations of WD 1856 b using the $16''$ telescope at Hereford Arizona Observatory, which is operated by Bruce Gary.  We observed WD 1856 b on 21 nights between May 18, 2020 and December 21, 2020 (in addition to the two HAO observations from October 2019 published by \citealt{Vanderburg2020}). The data collection and photometric analysis to produce light curves were performed as described by \citet{rappaport2016}. From the light curves, we derived transit times by fitting a simple model light curve to our observations. The transit dip is defined by five straight line segments: two steeply sloping segments at ingress/egress, two more gradual slopes nearing the transit center, and a flat bottom\footnote{\bedit{Even though WD 1856 b's transit does not have a flat bottom, this term is still useful in our modeling to precisely match the transit's shape. The running average applied to the transit shape effectively smears out the flat bottom to create a shape well matched to WD 1856 b's transit.}}. Six parameters define the transit model. Three of these parameters define the shape of the transit: the relative duration and depth of the steep segments and the more shallow segments, and the duration of the flat-bottom.  Three other parameters vary the overall depth, duration, and mid-transit time. After creating the piecewise linear transit model, we smooth that model with a running-average (boxcar) smoothing over 5 time bins, each 49 seconds long. While the resulting transit shape model does not perfectly describe the shape of high signal-to-noise observations like those from the GTC, it is certainly sufficient to describe the much noisier observations from the HAO $16''$ telescope.

\begin{figure*}
\centering
\includegraphics[width=\textwidth]{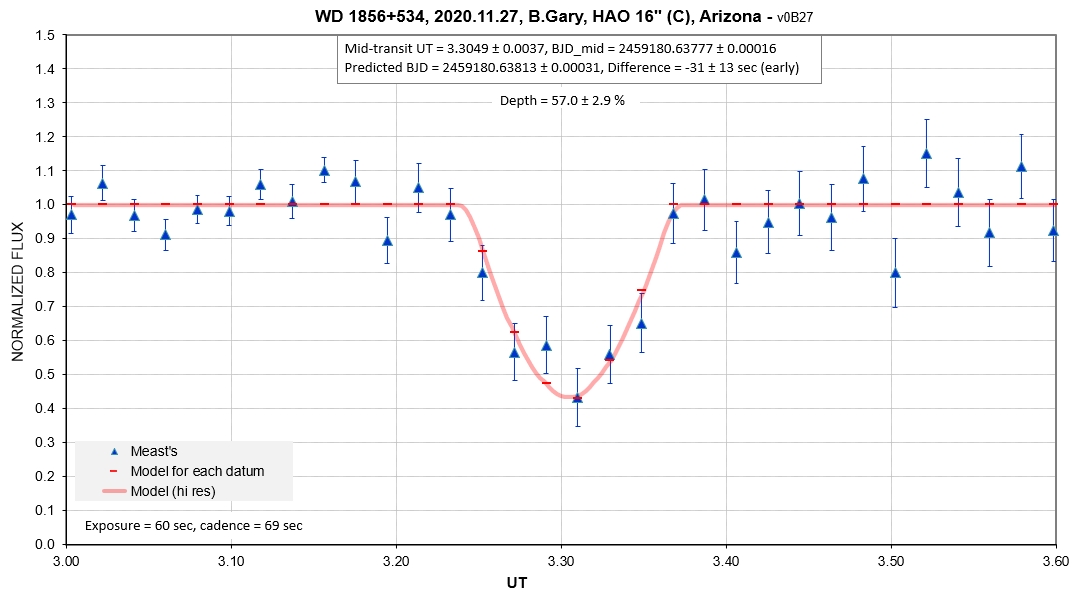}
\caption{ An example light curve obtained from the HAO on November \bedit{27}, 2020. Blue points are flux measurements of WD 1856, and the red curve is our model described in Section \ref{haodata}. We obtained 23 observations of WD 1856 b's transits with HAO and included them in our TTV analysis.  }\label{lightcurveex}
\end{figure*}



The high signal-to-noise GTC light curve from \citet{Vanderburg2020} was used for establishing the shape parameters of the transits (the relative depth and duration of the ingress/egress and more gradual slopes, and the duration of the flat bottom). Once we optimized the shape parameters based on the GTC light curve, we measured transit times by varying the overall depth, overall transit duration, and midpoint time of the individual transits. We performed the optimization using a spreadsheet that calculated the sum of $\chi^2$ between the observed and model light curves for many different parameter values and identifying the value that minimized $\chi^2$.  

We found no evidence for variations in any of the three parameters we optimized for each HAO observation. Unsurprisingly, the uncertainties on the transit times measured from the $16''$ HAO telescope are significantly (1-2 orders of magnitude) larger than the best uncertainties we measured with 10-meter class telescopes. However, the HAO observations could still be important because they have better time sampling, and can rule out large variations in the transit times when the oversubscribed 10 meter telescopes are unable to observe.

\section{Analysis}\label{Analysis}

\subsection{TTV Modeling}\label{stellarparameters}

We analyzed our transit times to assess the evidence for the presence of planets using a forward modeling approach. In brief, we used the \texttt{ttvfast} software \citep{Deck2014} to calculate the expected transit times for WD 1856 b in the presence of many different hypothetical planetary companions. We then compared the expected transit times for each planetary companion to the transit times we actually observed to determine the likelihood of each set of calculated transit times, and therefore identify which of the hypothetical orbits for planetary companions are preferred, plausible, or unlikely. 

The heart of our analysis is a likelihood function which, given a list of measured transit times and input planetary and orbital parameters for a hypothetical second planet in the WD 1856 system, returns the relative Bayesian likelihood of that particular system configuration. Our likelihood function, which we implemented in the Python programming language, performs the following steps:

\begin{enumerate}
\item First, the function identifies the ``transit number'' -- or the number of planetary orbits that had taken place since a reference transit, of each measured transit time. For our reference orbit, we took the transit observed by \citet{Vanderburg2020} with the GTC on October 22, 2019. We calculated the transit number of each time by calculating the time difference between each measured transit and the GTC observation, dividing by the orbital period reported by \citet{Vanderburg2020}, and rounding to the nearest integer. The transit number for each  observation we used is listed in Table \ref{transittimes}. 

\item We then calculated the expected transit times for WD 1856 b for a system \bedit{given arbitrary} input parameters using the \texttt{ttvfast} code \citep{Deck2014}. We initialized the model using the orbital parameters for WD 1856 b with an orbital period and time of transit as reported by \citet{Vanderburg2020}, an inclination and longitude of the ascending node of 0 degrees (defining the reference plane of the system to be the plane of WD 1856 b's orbit), and eccentricity of 0. \bedit{We chose to only explore circular orbits for WD 1856 b itself because the tidal circularization timescale for roughly Jupiter-mass companions (2 Myr) is much less than the age of the system (6 Gyr, \citealt{Vanderburg2020}), and because adding two additional variables to our parameter study would be prohibitively expensive.} \bedit{We define the reference frame for our calculations such that WD 1856 b's inclination is 90 degrees exactly, and we measure the inclination of the companion with respect to this value. We note that this value is slightly different from the actual inclination of WD 1856 b's orbit from the plane of the sky.} For these experiments, we assumed that WD 1856 b has a mass of 10 \mj, but because the planet's mass is much less than that of the star, this choice has little effect on our results. We allowed the planetary parameters of the hypothetical second planet in the system to vary.  We experimented with different time steps for the numerical integration and found that using a step of 0.02 days (1.5\% the orbital period of the innermost planet WD 1856 b) ensured that errors in the numerical integration were at least an order of magnitude smaller than the smallest uncertainties in our transit times. In each call to the likelihood function, we calculate the transit times of WD 1856 b over the course of 600 days, longer than the baseline \bedit{(431 days)} of our observations. 

Because \texttt{ttvfast} was originally designed to quickly model TTVs of low-mass planets near mean motion resonances, it makes two approximations that are not well justified for our experiments with the WD 1856 system. First, when \bedit{\texttt{ttvfast}} converts from orbital elements to positions and velocities, it assumes the planets are massless. Since our assumed mass for WD 1856 b of 10 \mj\ \bedit{is much smaller than the star’s mass (about 1.5\%) but not completely negligible}, this assumption causes the orbital period of long-period companions in the \texttt{ttvfast} simulations to differ slightly from their expected values. We worked around this issue by initializing the \texttt{ttvfast} simulations with a planetary mass for WD 1856 b of 0, and a stellar mass equal to the sum of its measured value (0.576 \msun, \citealt{Xu2021}) and our 10 \mj\ assumed value for WD 1856 b. The second approximation made by \texttt{ttvfast} is that it ignores the component of TTVs caused by the R\o mer delay, which is also known as the light travel time effect (K. Deck, private communication). In most known multi-planet systems, timing variations due to the R\o mer delay are negligible compared to TTVs caused by orbital variations, but in the case of a long-period companions orbiting a close inner pair like in the WD 1856 system, this effect can dominate (see, for example, Figure 7 of \citealt{rappaporttriples}). We therefore performed our own analytic calculation of the R\o mer delay (following Eqn 6 and 7 in \citealt{rappaporttriples}) and added the resulting curve to models returned by \texttt{ttvfast}. 

\item With these transit times from \texttt{ttvfast} (and corrected for the R\o mer delay) in hand, we performed some post-processing to adjust for slight differences between the orbital period we used to calculate the model transit times and the true orbital period of WD 1856 b. In systems with transit timing variations, the true mean orbital period of each planet can differ significantly from the instantaneous orbital periods that might be measured by observations over a short time baseline. To account for this, it is necessary to allow the true orbital period and transit time of the transiting planets to vary, often as free parameters in a Markov Chain Monte Carlo analysis. However, this would require adding two additional free parameters to our analysis, so we take a different approach to account for this effect.  We instead calculated a weighted linear least-squares fit between the transit times and transit number for both our observed transits and the model transit times from \texttt{ttvfast} where the weights were set equal to the inverse square of the uncertainties on the measured transit times. We then subtracted the best-fit line from both the observed and model transit times, so we could compare the residuals -- i.e. the TTVs themselves -- when calculating the likelihood of each model. Optimizing the mean orbital period and time of transit within each model calculation using this polynomial fit allows our model to efficiently tweak the model's assumed orbital period to match the observations without explicitly adding it as a free parameter. 

\item Finally, we calculated the likelihood of each model. We used a simple $\chi^2$ likelihood function where the likelihood $\mathcal{L}$ is given by: 
\begin{equation}
\mathcal{L} = e^{-\chi^2/2}
\end{equation}
\noindent and 
\begin{equation}
\chi^2 = \sum_i{\frac{(y_i - m_i)^2}{\sigma_i^2}}
\end{equation}
\noindent where $e$ is the base of the natural logarithm, $y_i$ are the individual transit timing variations (after subtracting the best-fit linear model), $\sigma_i$ are the uncertainties on each transit time, and $m_i$ are the model-predicted transit timing variations (also after subtracting the best-fit linear model). 

\end{enumerate}

\subsection{Parameter Study}\label{period_calculation}

With our likelihood function, we have the tool we need to assess the evidence for additional planets in the WD 1856 \bedit{system in hand}; we simply need to calculate the likelihood of many different orbital configurations for a hypothetical second planet in the system. Often, astronomers use sampling techniques like Markov Chain Monte Carlo to explore parameter spaces like this one, but because it is computationally expensive to calculate transit timing variations using numerical integration, and the space of possible companion orbits is very broad, we opt instead to explore slices in parameter space using grid searches. 

For our grid searches, we focused primarily on the planetary parameters that would likely change WD 1856 b's transit timing variations the most. In particular, we focused on the hypothetical second planet's mass, period, eccentricity, inclination, mean anomaly at the start of the integration, and argument of periastron. We treat two of these parameters, the mean anomaly and argument of periastron of the second planet's orbit, as nuisance parameters, and present our results in terms of the mass, period, eccentricity and inclination. 

\begin{figure*}
\centering
\includegraphics[width=\textwidth]{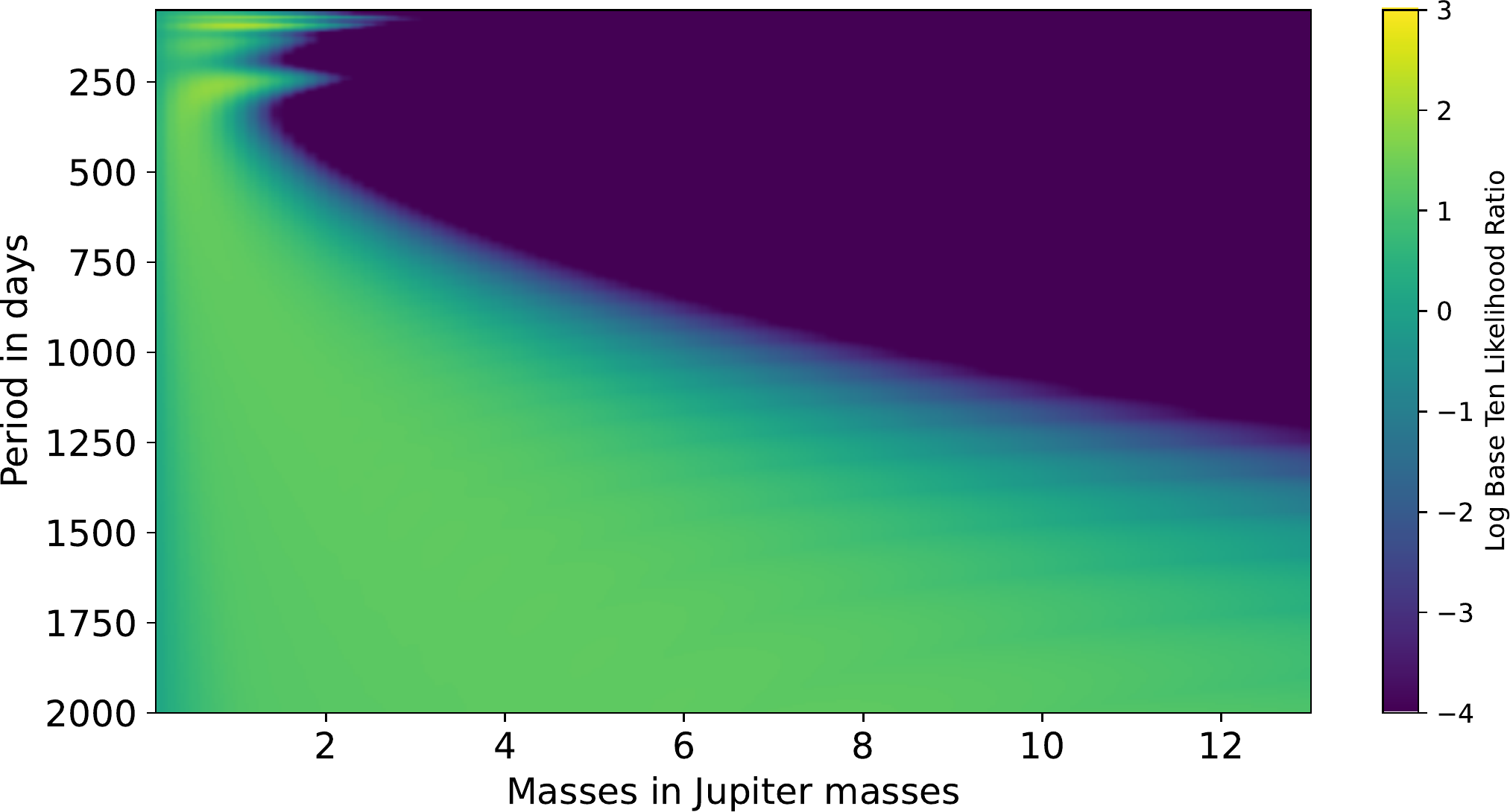}
\caption{ Relative likelihood of the presence of a companion planet to WD 1856 b as a function of its mass and orbital period, assuming a circular and coplanar orbit (inclination of 90$^\circ$ and eccentricity of 0). The color represents the logarithm (base 10) of the ratio of the likelihood calculated from the transit times assuming a companion planet and the likelihood assuming no additional planets. Purple regions of the plot show which parameter space is strongly ruled out by our observations, while we cannot necessarily exclude the presence of planets in the green and yellow regions.  Here we are looking at the parameter space we can rule out, purple area, based on the likelihoods of our transit timing variations.  In this particular slice of parameter space, we see that high mass planets and low periods can be ruled out, while low-mass and long-period planets generally cannot.}
\label{fig:singlemassperiod}
\end{figure*}

We created a three dimensional grid of different values of mass, period, and mean anomaly, and calculated the likelihood of a hypothetical companion to WD 1856 b with the parameters at each grid point.The grid had 300 mass points, 100 period points, and 100 mean anomaly points.  We collapsed the grid along the mean anomaly axis  by selecting the values for these nuisance parameters that maximized the likelihood, leaving grids showing only the most important physical parameters: mass/period, and eccentricity/inclination. As a point of comparison, we calculated the likelihood of a single-planet model (i.e. no TTVs). Figure \ref{fig:singlemassperiod} shows our results for a grid of outer companion orbits with different companion masses and orbital periods, with eccentricity fixed at 0, and the inclination of the outer planet fixed at 90 (perfectly co-planar\footnote{\bedit{Here again we define the reference frame of our calculations so that WD 1856 b's orbit has an inclination of 90$^\circ$ orbit. Because WD 1856 b's inclination with respect to the plane of the sky is about 88.8$^\circ$, our reference frame is slightly (1.2$^\circ$) different from WD 1856 b's actual orientation with respect to the plane of the sky.}} with the WD 1856 b).  The color represents the ratio between the maximum calculated likelihood within each grid point and the likelihood of a scenario with no additional planets (that is, no TTVs). In this case, we are able to rule out companions with large masses and short orbital periods in the purple region of the plot, while companions with low mass and long orbital periods are permitted, and in some cases, even weakly favored \bedit{(although not significantly; see Section \ref{noplanets}).} 

To explore the relationships between the mass, orbital period, and two additional parameters (eccentricity and inclination of the companion's orbit), we generated more grids slicing through a six-dimensional likelihood space. These will be available upon publication.  We repeated our calculations to produce grids of the relative likelihood of different values of mass/period and eccentricity/inclination while fixing the other parameters to different values. In cases where we explored circular orbits only, we used three-dimensional grids as described in the previous paragraph, but when we explored non-zero values for eccentricity, we calculated the likelihood function along a four dimensional grid that also included the argument of periastron for the outer companion's orbit\footnote{We only included argument of periastron as a grid parameter when eccentricity was nonzero because the argument of periastron and the mean anomaly are perfectly degenerate when eccentricity is 0.}. In cases where eccentricity was non-zero, we collapsed our results into a two-dimensional grid by maximizing the likelihood over both the mean anomaly and argument of periastron. These additional grid calculations give us a better understanding of the impact the other variables had on the final likelihood and the effect different combinations of these parameters have on which planetary orbits are plausible and which can be ruled out. 

We discuss the results of these calculations in Section \ref{planetlimits}. We performed these computationally intensive calculations using the MIT SuperCloud computer cluster \citep{reuther2018interactive}. Additionally, we attempted to reproduce the results from our  numerical experiments using \texttt{TTVfast} with an analytic approach. We used the expressions for the physical and R\o mer delays from \citet{rappaporttriples} to estimate TTV amplitudes over the baseline of our observations for similar grids of parameters. While the analytic expressions from \citet{rappaporttriples} are approximate and ignore higher order terms, they are very computionally efficient to calculate. In general, we found good agreement between the numerical and analytic approaches for relatively low mutual inclinations and eccentricities, but at higher inclinations and eccentricities, the analytic approximations were less accurate. The analytic results gave good qualitative agreement with the numerical calculations, increasing our confidence in these results.

\section{Results} \label{Results}

\subsection{No Evidence for Additional Planets Orbiting WD 1856}\label{noplanets}

The first result of our analysis is that we found no evidence for transit timing variations, and therefore no evidence for additional planets orbiting WD 1856 b, at least to the sensitivity limits of our dataset. We identified the model with the highest likelihood from our entire parameter study and compared it to the likelihood of a model without a second planet.\bedit{ We consider parameter space ruled out when the likelihood is a factor of $10^{4}$ times worse than a 1 planet model}. Although the likelihood of the best-fit model from our entire study was higher than the likelihood of a single-planet model, this could be because a two-planet model has significantly more degrees of freedom than a single-planet model. To account for the increased model flexibility, we calculated the Bayesian Information Criteria (BIC) to determine whether the higher likelihood of our best two-planet model than a single planet model is strong evidence for the existence of another planet. The BIC is an approximate method for comparing two models; the model with the lowest BIC is favored, and one model is only strongly preferred over another when the difference in BIC (or $\Delta$BIC) is more than 10.  We found for the single-planet scenario the BIC was 64.2, while for the best two-planet model, the BIC was 78.1. Therefore, the single-planet model is strongly favored over even our best two-planet model, so we find no convincing evidence of another planet in the system. \bedit{We note that the parameters for the best two-planet model were 2 Jupiter masses, 50 days, 0.284 eccentricity, 70.5$^\circ$ inclination,  301.2$^\circ$ argument of periastron, and 88.2$^\circ$ mean anomaly.}

In the absence of TTVs from additional planets, we can measure the best-fit orbital period and time of transit for WD 1856 b. We performed a linear fit to the transit times (allowing for some excess noise in the transit time measurements) using a Differential Evolution Markov Chain Monte Carlo algorithm \citep{terbraak}. We measured an orbital period of $1.407939217 \pm  0.000000016$ days and a time of mid-transit of $2459038.4358981 \pm 0.00000114$, in BJD, chosen at the epoch which minimizes the covariance between the period and transit time.\bedit{We summarize this information in Table \ref{transitorbitalperiod} as well.} This ephemeris has high enough precision to predict transit times with an uncertainty better than 40 seconds for the next century. 

\begin{table}
	\begin{center}
	\caption{\bedit{Best Fit Orbital Period and Time of Transit for WD 1856 b.}}
\label{transitorbitalperiod}
\scriptsize
	\begin{tabular}{llcll} 
		\hline
		Parameter & Value & &  Uncertainty & Unit\\
		\hline
\textit{Period} & 1.407939217  & $\pm$ & 0.000000016 & days\\ 
\textit{$t_0$ (at epoch 0)} & 2458779.3750822 & $\pm$ & 0.00000316 & BJD$\_$TDB \\
\textit{$t_0$ (minimal covariance)} & 2459038.4358981 & $\pm$ & 0.00000114 & BJD$\_$TDB \\
		\hline
	\end{tabular}
	\end{center}
\end{table}

\subsection{Limits on the Presence of Additional Planets Orbiting WD 1856}\label{planetlimits}

Although we find no affirmative evidence for additional planets in the WD 1856 b system with the current data, it is still possible that such planets could exist but that the current data is not sensitive to them. To determine the parameter space in which planetary companions have and have not been solidly excluded based on the current data, we constructed portraits of the computed model likelihood as it depends on the orbital parameters of the additional planet. 
We examined the likelihood for various sets of orbital parameters (the companion's orbital period $P_c$, eccentricity $e_c$, and inclination $i_c$) and physical parameters (companion planet mass $m_{c}$) by constructing parameter grids and computing the model likelihood for a discrete set of companion parameters. 

In the portraits that follow, in general regions of low probability (which tend to have shorter orbital period and larger companion planet mass) are ruled out because they would produce TTVs significantly larger than the observed values. On the other hand, the current data do not allow us to rule out companions in all of the studied parameter space; companions whose likelihoods are commensurate with (or even slightly higher than) the raw likelihood of the single-planet model cannot be excluded with the present data, even though the BIC \bedit{indicates the data quality does not support the added complexity of the two-planet solution generally. } 

\begin{figure*}
\centering
\includegraphics[width=\textwidth]{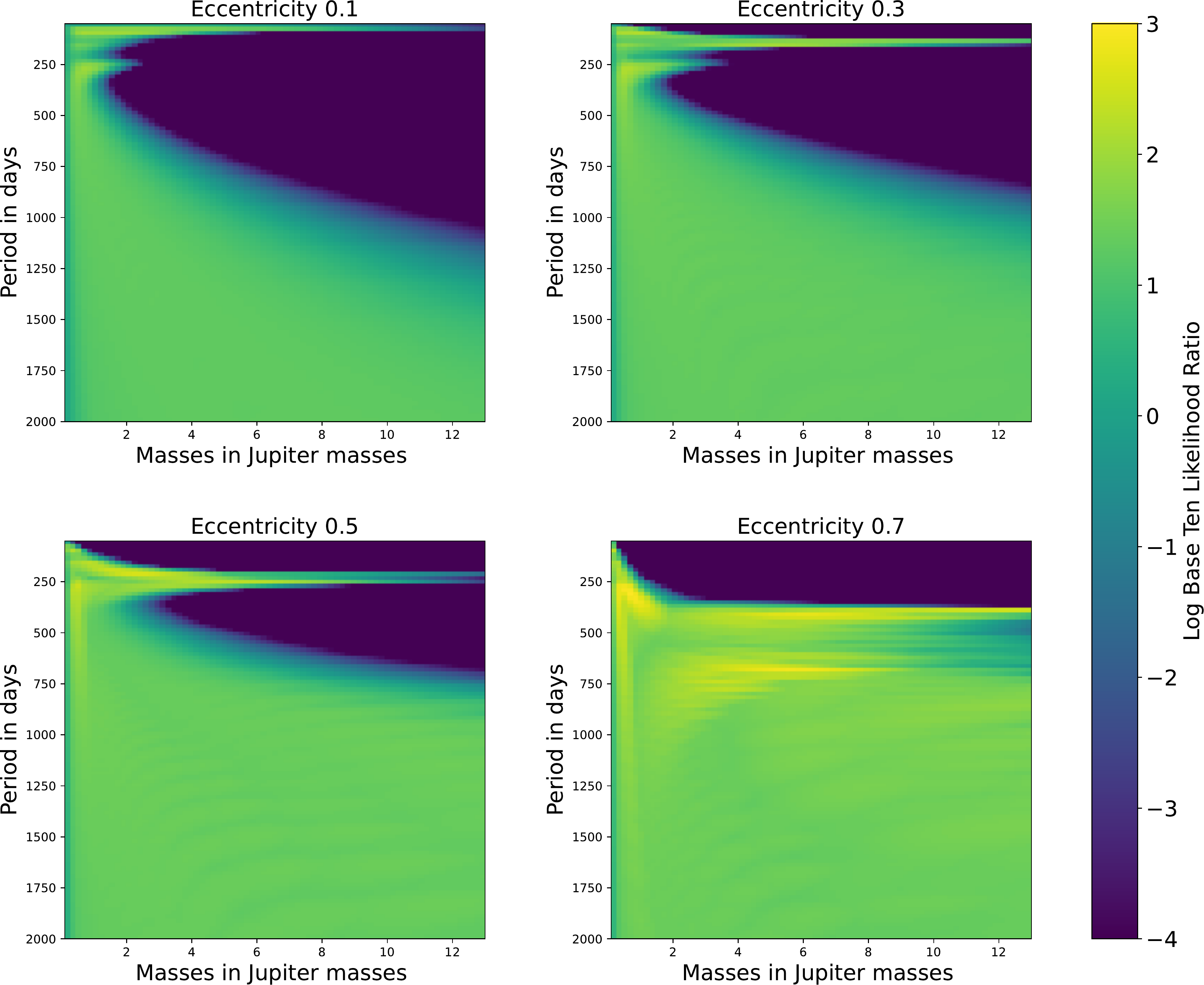}
\caption{ The parameter study of a handful of fixed values of eccentricity and changing masses and periods, with inclination fixed at mutual inclination angle = 0. As in Figure \ref{fig:singlemassperiod}, purple regions of parameter space are strongly ruled out, while companions may exist in the green/yellow regions of parameter space.  We again see in these slices that larger companion planet masses and shorter companion orbital periods are disfavored, although our constraints are generally weaker at high eccentricity. We also see spikes where high-mass companions are allowed at particular short periods at moderate to high eccentricities that we think are due to poor sampling of transit times at those periods and occasional outliers in our data.  }
\label{fig:eccstudy}
\end{figure*}

\subsubsection{The Effect of Orbital Eccentricity}
In Figure \ref{fig:eccstudy}, we show the we show the result of one such analysis, where the likelihood is computed as a function of $m_{c}$ and $P_c$ for four slices of companion planet eccentricity $e_c$. 
The color represents the ratio between the likelihood of a model at each grid point in mass/period and the likelihood of a model with no additional planets. The general trend persists that larger companion planet masses and shorter companion orbital periods (which together would tend to exert larger TTVs on WD 1856 b) are disfavored. 
Within the most favorable area of parameter space (lower $m_c$ and higher $P_c$), an additional band of apparent increased likelihood emerges for the planetary companions that induce small but non-zero TTVs to match the fluctuations in the data. Higher companion eccentricity results in a smaller range of possible orbital periods, as a closer periastron results in more significant planet-planet interactions (and \bedit{larger} TTVs).

In addition to those general trends with eccentricity, a secondary structure in likelihood emerges, which takes the form of horizontal spikes of allowed regions at particular orbital periods, and in some cases extending to high masses. We think that these spikes are caused by a combination of poor sampling of transit times at those periods, and occasional outliers in our transit timing data. In regard to sampling these spikes appear at difficult periods for us to measure. We often see them at periods close to 1 year or 1/2 year since we are unable to observe WD 1856 b when its location in the sky is too close to our Sun, and the star is only overhead during daytime. This causes gaps in our observations each year which prevent us from ruling out large timing variations at those times.  We find that the spikes in the likelihood graphs are more prominent at high eccentricity because higher eccentricity introduces faster changes into the TTV curve, so it is possible that one or two outlying points can strongly affect the likelihood and favor a fast change to the TTV curve.

\begin{figure*}
\centering
\includegraphics[width=\textwidth]{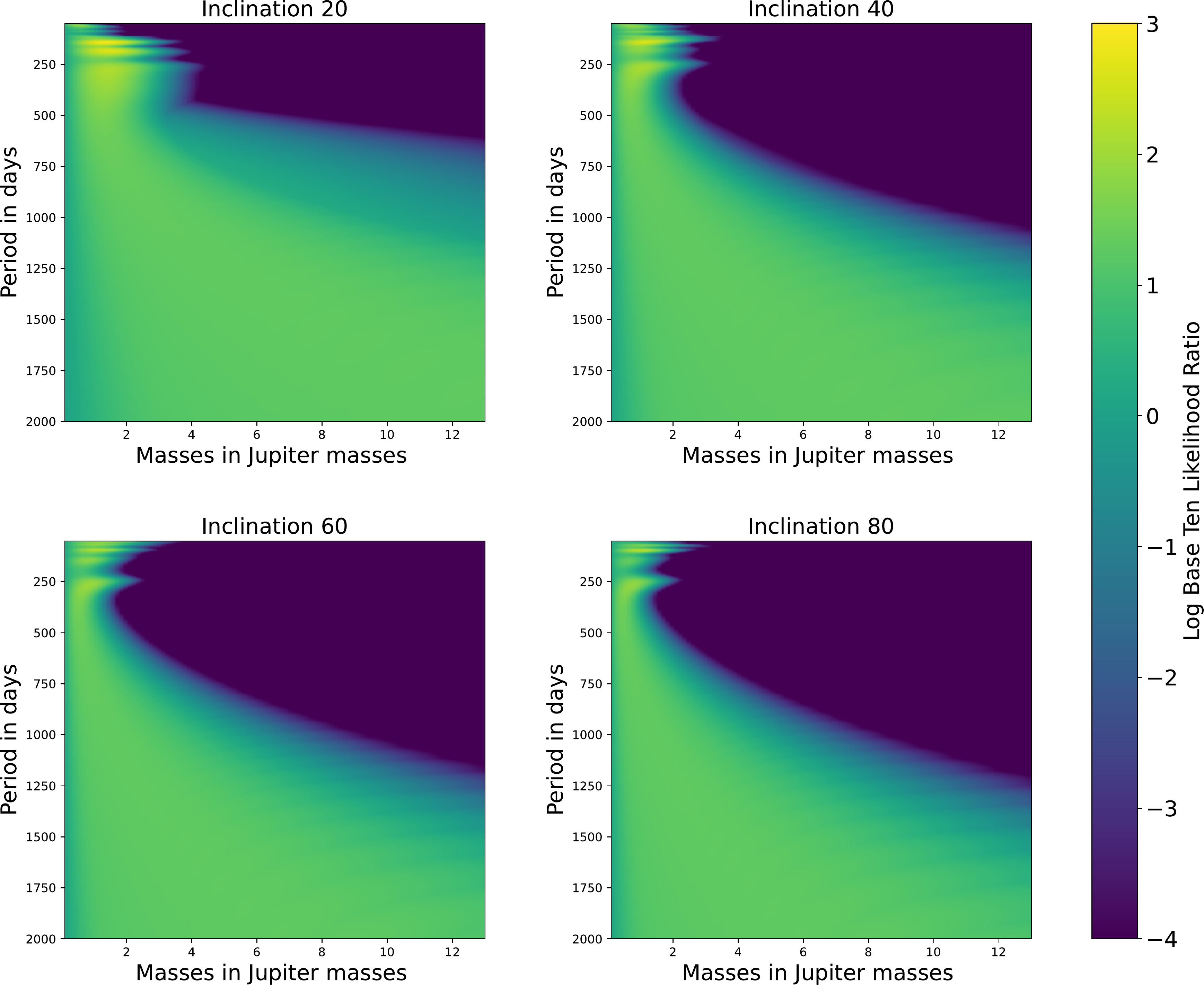}
\caption{The parameter study of a handful of fixed values of inclination and changing mass and period, with eccentricity fixed at 0. Again, the color scheme is the same as described in Figure \ref{fig:singlemassperiod}, where we rule out companions in the purple regions.  We see the same pattern here as in Figures \ref{fig:singlemassperiod} and \ref{fig:eccstudy}, where large companion planet masses and shorter companion orbital periods are disfavored in these slices. We see the strongest constraints at long-periods with low mutual inclination (companion inclination near 90$^\circ$), but even at high mutual inclination, we rule out are able to rule out lower mass and shorter period companion planets. At low mutual inclination (near 90$^\circ$ companion inclination) we primarily rule out companions based on contributions to the R\o mer delay, and at high mutual inclination, we instead primarily rule out companions thanks to the physical delay.}
\label{fig:incstudy}
\end{figure*}


\subsubsection{The Effect of Orbital Inclination}

In Figure \ref{fig:incstudy}, we show  the result of another analysis, where the likelihood is computed as a function of $m_{c}$ and $P_c$ for four slices of companion planet inclination $i_c$. The general trend persists that larger companion planet masses and shorter companion orbital periods are disfavored in these inclination slices. In Figure \ref{fig:incstudy}, we label the inclination of the companion with respect to the plane of the sky -- that is, face-on orbits have an inclination of \bedit{0}. Since \bedit{in our reference frame} the inner planet, WD 1856 b, is transiting with an inclination of 90 degrees, and we set the longitudes of ascending nodes for the two planets to be equal, the mutual inclination between the two orbits is $90-i_c$.   

In general, we find that we are able to rule out lower-mass and shorter-period companion planets when the mutual inclination between the planets is low. The data allow highly inclined companions with relatively large masses and short orbital periods.  We are able to rule out more parameter space when the companion's inclination is near 90 degrees (low mutual inclination) because of the R\o mer delay, which is proportional to $sin(i_c)$. For inclinations near 0 we see additional effects at shorter periods. These effects are caused by the physical delay, or actual changes in the orbit of the potential planet. Unlike the R\o mer delay, the amplitude of the physical delay is actually larger at high mutual inclinations (see Eqns 8-10 of \citealt{rappaporttriples}). 

\begin{figure*}
\centering
\includegraphics[width=\textwidth]{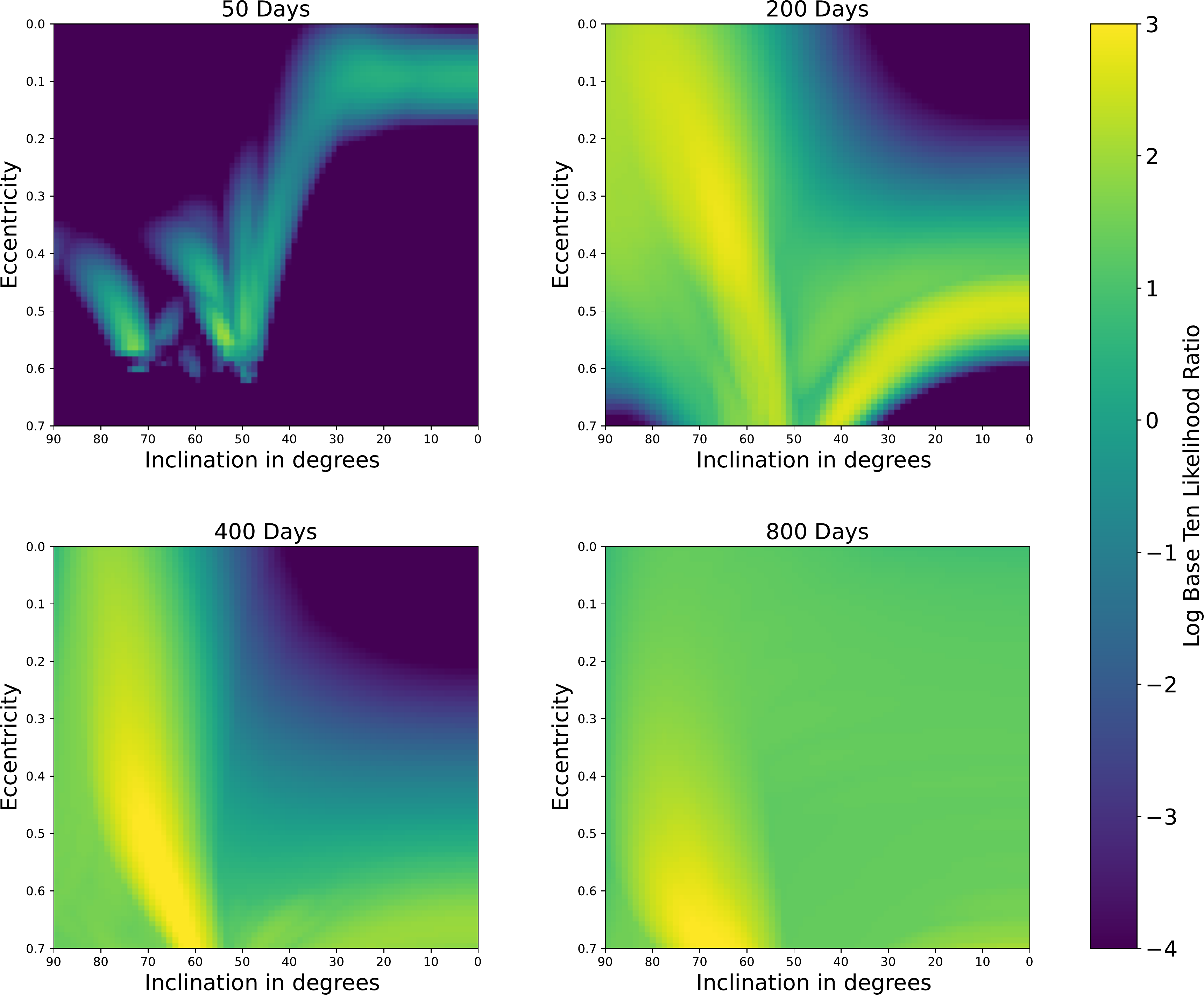}
\caption{ The parameter study of changing eccentricity and inclination, with the orbital period fixed to a handful of different values and the with mass fixed at 2 \mj. In general, as the orbital period increases, we are able to rule out less parameter space. At 200 days there is a cross over where the R\o mer delay dominates over the physical delay, which appears to cause a qualitative change in the pattern of which parts of parameter space are favored/disfavored. At higher periods like 400 and 800 days we can rule out little parameter space because those periods are longer than our observing baseline.}
\label{fig:periodstudy}
\end{figure*}

\subsubsection{The Effect of Orbital Period on an Eccentricity/Inclination slice}

In Figure \ref{fig:periodstudy} we show the result of an analysis wherein the likelihood is computed as a function of the $e_c$ and $i_c$ for four slices of companion planet period in days $P_c$. \bedit{In this analysis the mass of the planet in our parameter search was set to 2 $M_J$.}  In general as we increase the period of our parameter slices we are able to rule out less parameter space. 

At 50 day periods, we are able to completely rule out companions in a significant fraction of parameter space. This causes a particular pattern to emerge in the likelihood slice, where only very particular combinations of eccentricity and inclination can produce TTVs as small as we measure. At short periods, the TTV curves are dominated by the physical delay, which produces large TTVs at high mutual inclination and high eccentricity. We also see the contribution of the R\o mer delay at low mutual inclination, but this effect is subdominant (see Figure 7 of \citealt{rappaporttriples}). 

At 200 day periods we see a crossover to where the R\o mer delay dominates over the physical delay. This fundamentally changes the pattern of which eccentricities and inclinations are allowed or ruled out. Even though the R\o mer delay is strongest at low mutual inclination, we find that we cannot rule out this space when the eccentricity of the planet is low. We can, however rule out parameter space with high eccentricity and low mutual inclination, because as the eccentricity increases, the amplitude of both the R\o mer delay and physical delay increase.  At even longer periods, ranging from 400 days to 800 days, we are able to rule out relatively little parameter space for planets of this mass (2 \mj). 





\begin{figure*}
\centering
\includegraphics[width=\textwidth]{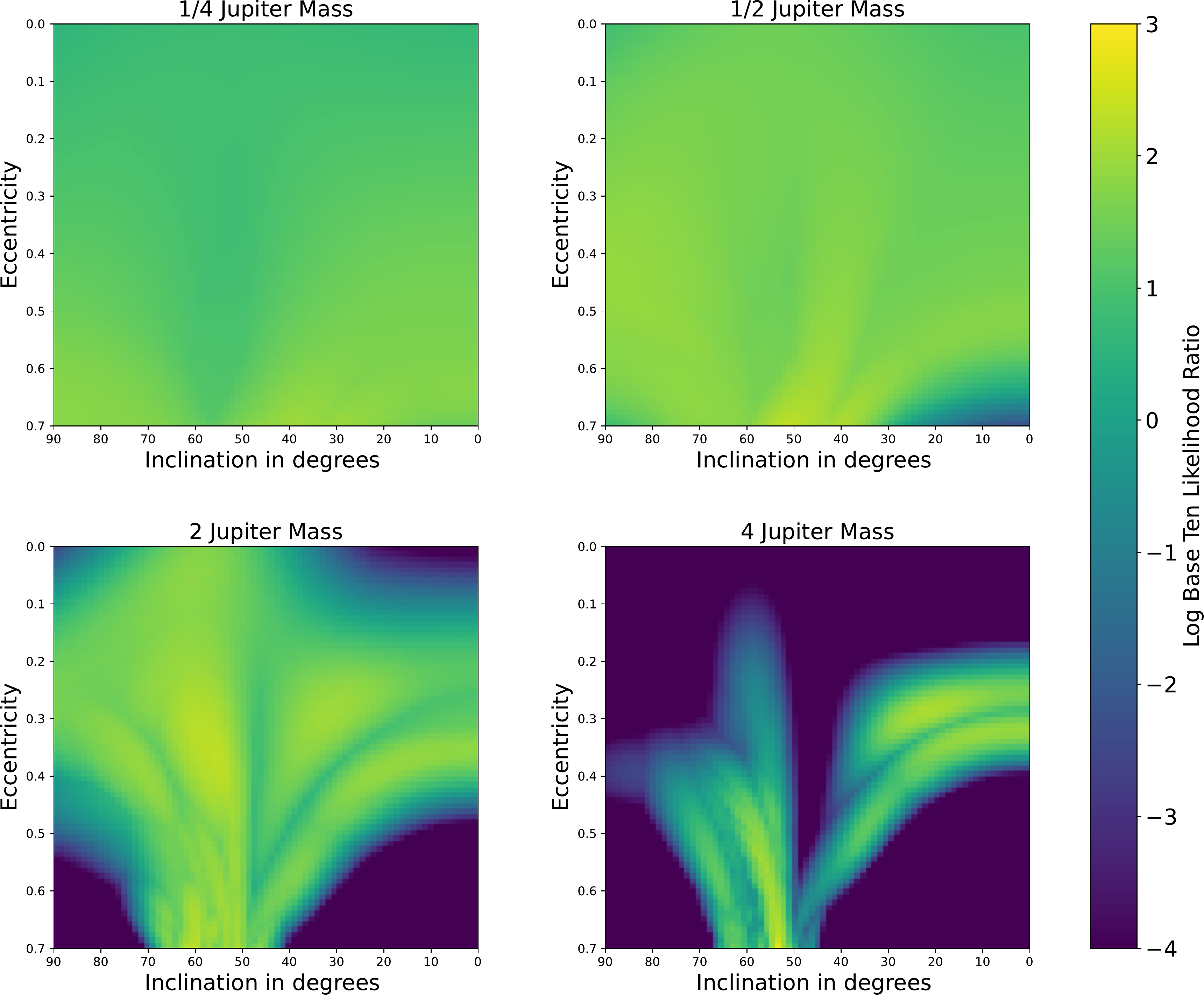}
\caption{The parameter study of a handful of fixed values for mass and changing eccentricity and inclination, with the orbital period fixed to 130 days. In our analysis we see that low-mass planets can exist at almost any orbit, but we are able to rule out planets with larger masses in wide ranges of parameter space. Because the amplitude of both the physical and R\o mer delay scale linearly with mass, the pattern of which regions of parameter space are favored/disfavored remain relatively similar within the slices.}
\label{fig:massstudy}
\end{figure*}

\subsubsection{The Effect of Mass on an Eccentricity/Inclination slice}

Finally, in Figure \ref{fig:massstudy} we show the result of an analysis where we calculate likelihood as a function of eccentricity and inclination of the companion's orbit, while holding the orbital period fixed at 130 days, and using various values of the planet's mass. Because the mass of the planet linearly scales both the TTV amplitude from the R\o mer delay and the physical delay, the pattern we see (in which parts of parameter space are ruled out and which are allowed) in the likelihood slices does not change significantly at the different masses. Instead, changing the mass mostly affects the ``stretch'' of the images and enhances or decreases the contrast between the regions. In general we find that for low planet masses, $m_{c}$, the companion can exist in almost any orbit, and larger $m_{c}$ rule out more parameter space with these specific configurations. Near 90 degrees, companions are largely ruled out is due to the R\o mer delay, since that was where it has the strongest effects. On the other hand, near 0 degrees we rule out parameter space because of the physical delay, which is due to the prominence of this effect at lower inclinations (higher mutual inclinations). At middling inclinations around 45$^\circ$, both of these effects are attenuated, so we can rule out relatively few companion orbits in this region of parameter space.

\section{Discussion and Conclusions} \label{discussion}

 

\subsection{Implications for WD 1856 b's formation}

The main goal of our study was to determine whether additional planets, beyond WD 1856 b, reside in the WD 1856 system, and subsequently to determine whether they may have had a role in WD 1856 b's migration. If additional planets orbit WD 1856 at farther distances from the star, they could plausibly have acted as dynamical perturbers to send WD 1856 b into its current short-period orbit. However, we found no evidence for such planets in the transit time measurements that we analyzed. Although there were some combinations of parameters that produced TTV curves that yielded slightly higher likelihoods than a model with no additional planets, these models were not favored in a comparison of the BIC for the two scenarios.  We therefore find no evidence supporting the hypothesis that WD 1856 b migrated to its current orbit as the result of dynamical interactions with additional planets in the system.  

However, even though we find no evidence for additional planets orbiting WD 1856, this does not rule out dynamical interactions with other planets as the cause of WD 1856 b's inward migration. Our data were collected over the course of just two observing seasons, and even in the most favorable circumstances, we can only rule out planetary mass companions at periods less than about 1200 days (or 4 years). Any planets massive enough to dynamically perturb WD 1856 b into a high eccentricity orbit  must have originally orbited at distances greater than 2-4 AU in order to survive post-main-sequence evolution, and therefore have orbital periods of at least a decade \citep{nordhaus}. Therefore, even if WD 1856 b did migrate to its current location as the result of dynamical interactions with additional planets in the system, we would not expect to detect them given the current limits of our observations. If we find no evidence for additional planets with transit timing measurements, there may be other way to detect other planets in the WD 1856 b system including astrometry \citep{Sanderson2022}, direct imaging \citep{Xu2015}, and thermal emission \citep{Limbach2022}. 

While we currently find no evidence of additional planets in the WD 1856  system, more data could tell a different story. Additional transit observations with large telescopes would allow updates to the likelihood maps presented in this work and increase our sensitivity to long-period planets.   Occasional observations over the course of the next 5-10 years would likely extend our sensitivity to planetary companions orbiting where we expect dynamical perturbers of WD 1856 b might reside. So while our findings can eliminate some parameter spaces, more transit data are needed to fully understand the geometry and population of the WD 1856 system and identify any possible neighbors that it has that could explain the movement of WD 1856 b. 



\begin{figure*}[h!]
\centering
\includegraphics[width=0.9\textwidth]{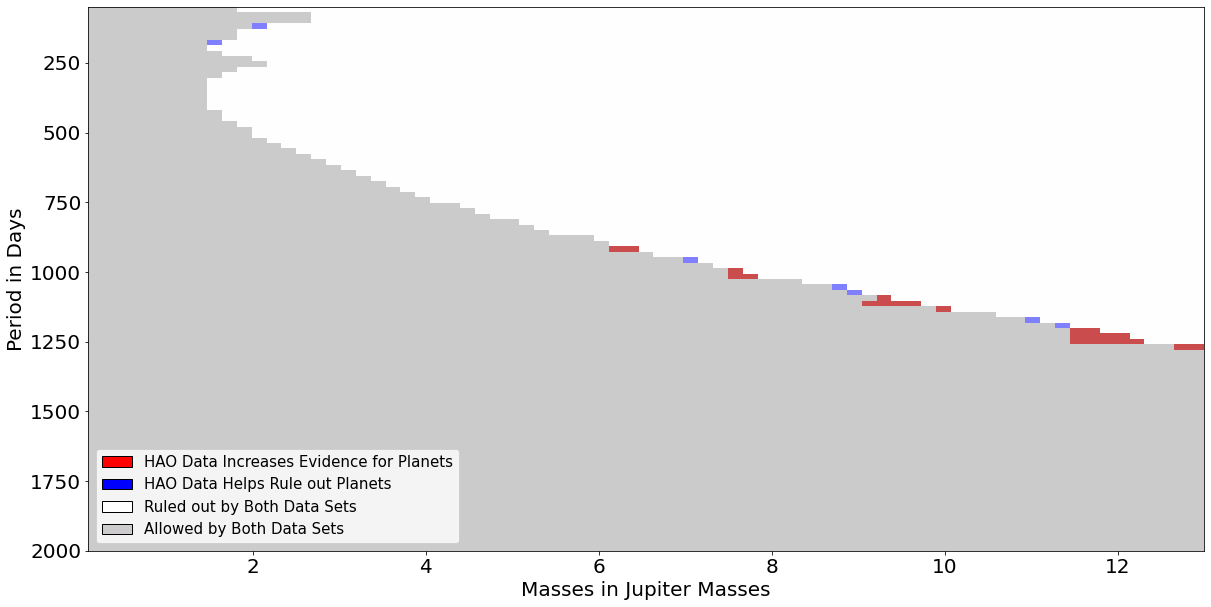}
\includegraphics[width=0.9\textwidth]{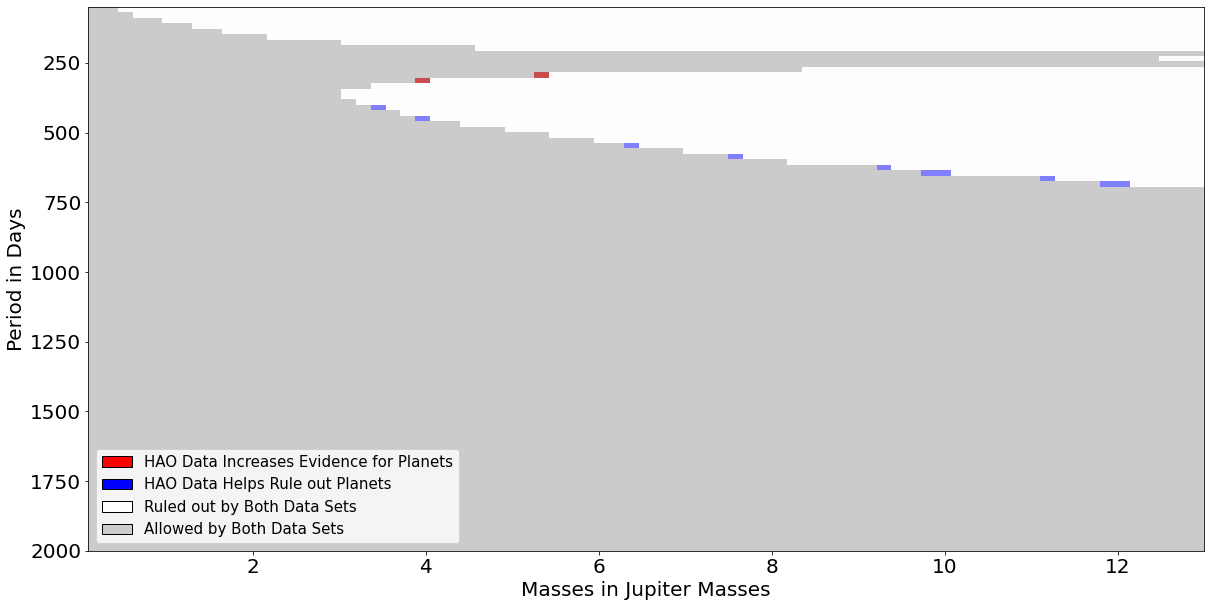}
\caption{ A visualization of how data from small telescopes helps in ruling out parameter space. Each panel color-codes different regions of parameter space based on whether adding or removing data from small telescopes changes whether hypothetical companion planets in those regions of parameter space are ruled out (defined as having a ratio between the likelihood for a model including third planet to a model without a third planet less than $10^{-4}$). Regions ruled out in both cases are grey, regions where adding data from small telescopes help rule out more space are blue, and regions where adding data from small telescopes rules out less parameter space are red. Top: A comparison in a parameter study of varying masses and periods with eccentricity fixed to 0 and an inclination fixed to 90 degrees. Bottom: A comparison in a parameter study of varying masses and periods with eccentricity of 0.5 and an inclination of 90 degrees. For circular orbits, we find little evidence that including data from small telescopes improves our constraints, at higher eccentricity, the improved sampling possible with small telescopes can help improve constraints, even if on the margins. }
\label{bigvssmall}
\end{figure*}

\subsection{The value of transit times from small telescopes}

One novel component of our work is that we combine transit times measured from small ($\lesssim$ 1-m diameter) telescopes with those measured from large ($\gtrsim$ 8-m diameter), professional observatories. Although the observations from large, professional telescopes yield much more precise transit time measurements than smaller observatories, it is more difficult to secure observing time on these facilities, and relatively few observations are possible. Given these trade-offs, we investigated whether data from the small telescopes significantly improved our constraints compared to only using observations from large professional telescopes. 

To test how much value is added by the inclusion of transit times from small, privately owned telescopes, we performed a comparison of how much parameter space we were able to rule out when using our full transit timing data set as opposed to only the sparse, but precise, timing observations from 8-10m class telescopes. We repeated our analysis for two likelihood slices in mass/period space (with coplanar orbits, and eccentricity fixed at 0 and 0.5) using only data from 8-10m class telescopes, and compared them with our nominal analyses using our full dataset. Figure \ref{bigvssmall} summarizes the differences between the two analyses. The figure highlights regions of parameter space where the addition of data from small telescopes either marginally increases or marginally decreases the likelihood ratio between a multi-planet model and a single-planet model across a threshold of $10^{-4}$.  In general, the constraints obtained in the two analyses are very similar, with only slight differences on the margins (where the small telescopes may help provide slightly stronger constraints for planets with eccentric orbits). This suggests that the Gemini and GTC data dominate the combined data set, and transit observations from small telescopes do not contribute significantly.

\begin{figure*}
\centering
\includegraphics[width=0.9\textwidth]{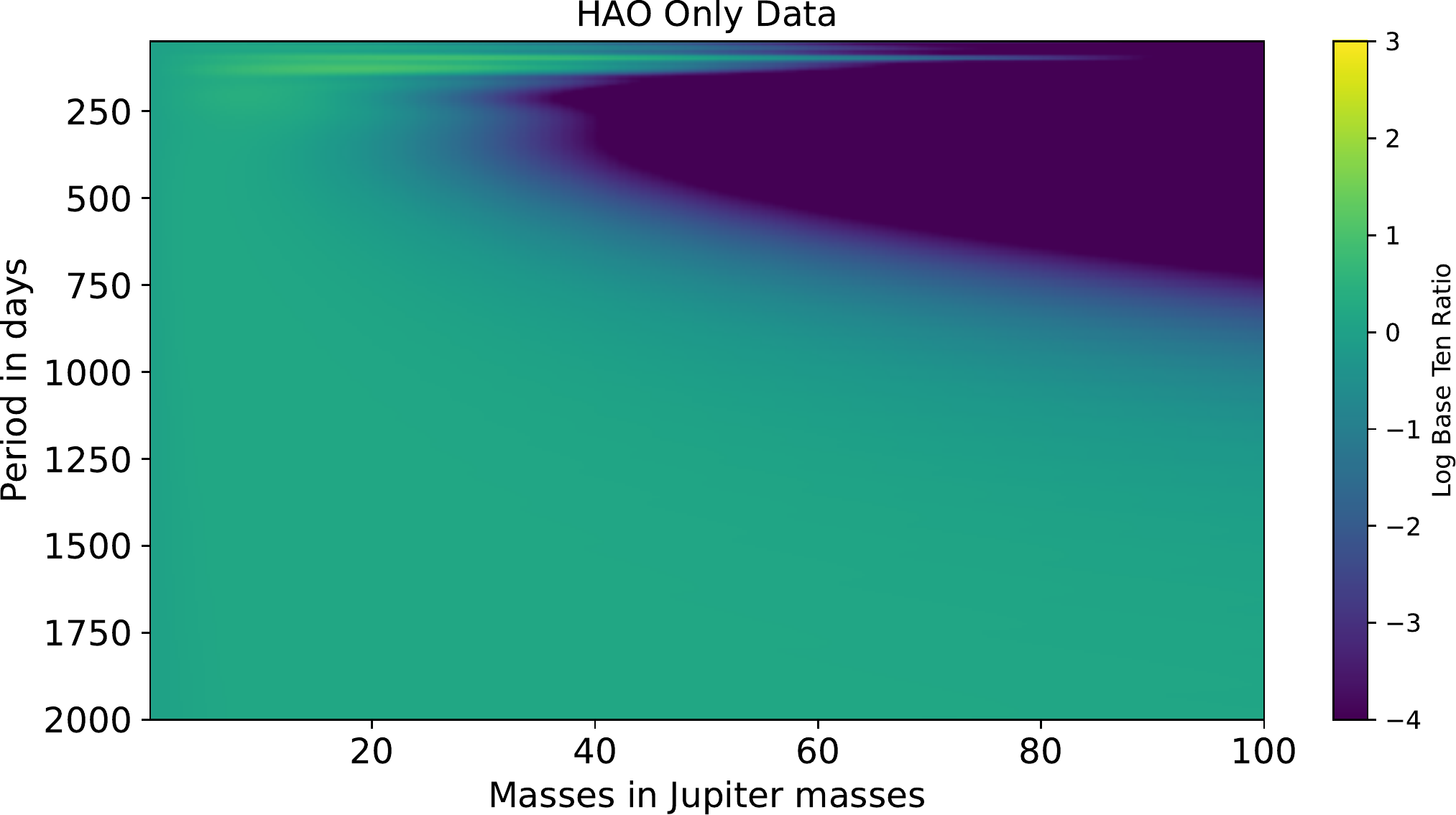}
\includegraphics[width=0.9\textwidth]{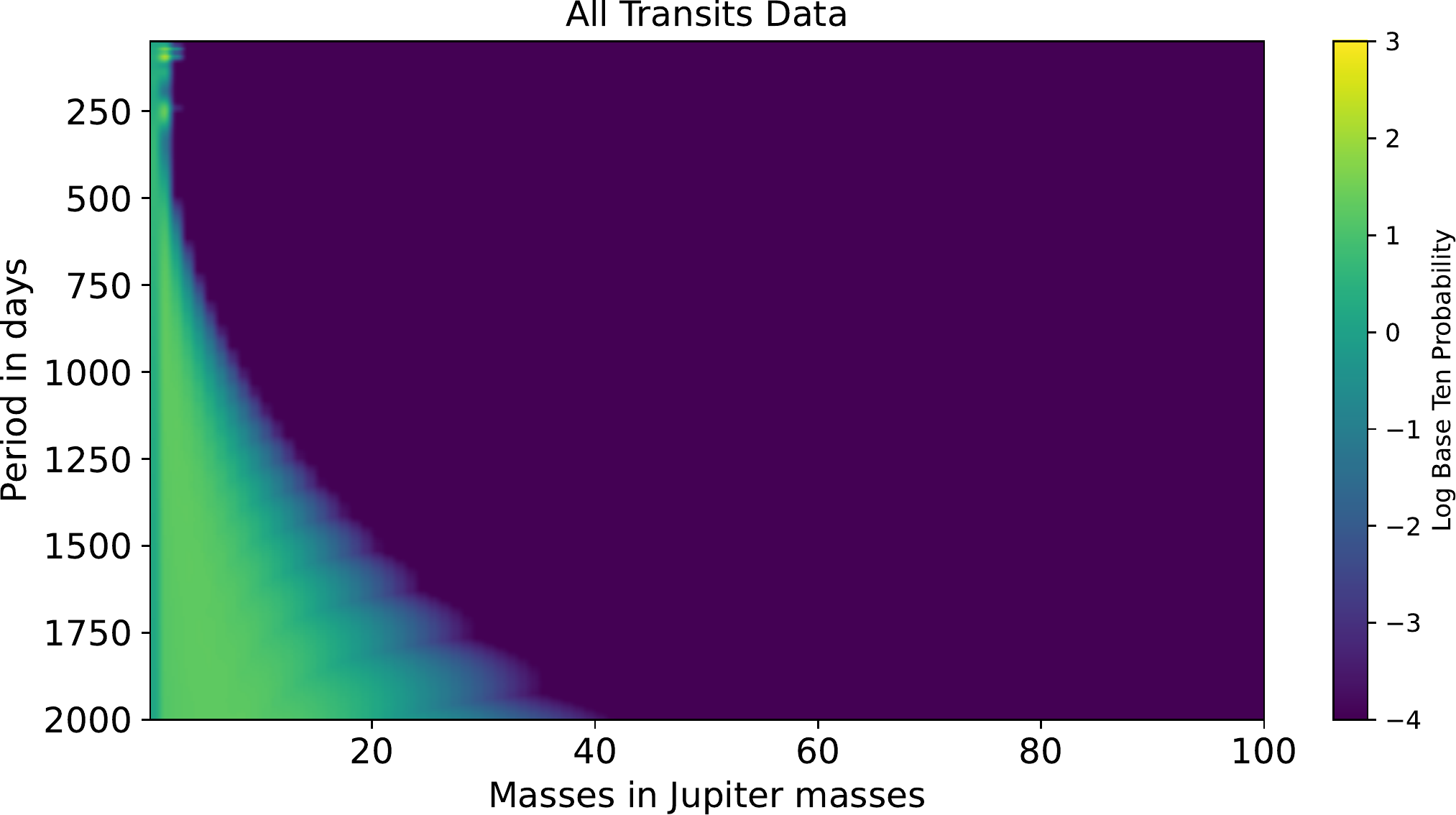}
\caption{ Relative likelihood of the presence of a companion planet to WD 1856 b as a function of its mass and orbital period, assuming a circular and coplanar orbit (inclination of 90$^\circ$ and eccentricity of 0),  with only data from HAO in the upper graph. This figure is analogous to Figure \ref{fig:singlemassperiod}, which covers a narrower range of masses using the full set of observations but uses the same color scale and orbital period axis.\bedit{This lower figure is analogous to Figure \ref{fig:singlemassperiod}, and uses the same color scale and range of orbital periods, while covering a much wider range of masses.} We find that data from small observatories like HAO could identify companions in the brown dwarf mass range, but probing Jupiter-mass planets requires more precise transit times from larger (8-10m class) telescopes. }
\label{haoonly}
\end{figure*}

We also compared the constraints we were able to obtain when combining the dataset with only the less precise transit times from HAO. We repeated our analysis for one likelihood slice in mass/period space, with coplanar circular orbits for the companion. Figure \ref{haoonly} shows the results of this analysis. Qualitatively, the parameter space ruled out by the HAO observations is similar to that ruled out by the full dataset, with strong constraints on the presence of short-period massive companions, and weaker constraints for low-mass companions at longer orbital periods. However, the lower precision on the transit times measured by HAO shifts the constraints towards higher masses, and in the best cases HAO observations can only rule out masses larger than about 30 \mj. Evidently, HAO data alone are not competitive with the sparse but precise timing measurements from 8-10m class telescopes.

\subsection{Prospects for Future Observations}

In this work, we demonstrated that it is possible to measure transit times of WD 1856 b with sufficient precision to detect \bedit{(or rule out)} distant \bedit{super-Jovian} planets in the system. Currently, our observational baseline is too short to detect planets with the expected orbital periods for planets that would have survived WD 1856's red giant phase, but future observations will increase our sensitivity to long-period planets. It would be highly beneficial to collect more transits from the WD 1856 system over the coming years to extend our sensitivity to detect or rule out longer-period planets. 


Some highly precise transit observations are already planned; the \textit{James Webb Space Telescope} (JWST) will observe multiple transits of WD 1856 b during its first observing cycle \citep{macdonaldjwst,vanderburgjwst}. Although these observations are aimed and studying the planet's atmosphere, they will likely yield highly precise transit time measurements, with uncertainties similar to the GTC and Gemini observations taken from the ground. The JWST data will extend the transit timing baseline with large, professional telescopes to more than 3 years. Our current observational baseline of about 300 days gives sensitivity to planets at orbital periods up to 1200 days, so increasing our baseline to over 1300 days could give us sensitivity to planets with periods of up to a decade. Over that time, it likely will not be necessary to maintain the same high cadence of transit observations we present here, since we already strongly constrain short-period planets. Instead, we may be able to increase the spacing of observations exponentially over time as we probe longer orbital periods. 


\section{Conclusions}

In this work, we have presented the first constraints on additional planets orbiting WD 1856+534 from transit timing measurements. Our conclusions are summarized as follows: 

\begin{enumerate}
    \item We find no compelling evidence for additional planets orbiting WD 1856. Our observations are sensitive to \bedit{2} Jupiter-mass planets in orbits out to 500 day periods, and 10 Jupiter-mass planets in periods as long as 1000 days. 
    \item We are most sensitive to planets with short orbital periods and high masses, as expected from analytic arguments \citep[e.g.][]{rappaporttriples}. Our constraints depend in more complex ways on the orbital eccentricity and inclination of hypothetical companion planets.
    \item We find that our constraints are dominated by highly precise transit timing measurements from 8-10 meter class telescopes, and even relatively few and sparsely sampled observations provide much stronger constraints than larger numbers of more frequent but less precise transit measurements from small telescopes. 
    \item Our constraints on the existence of additional planets do not yet extend to long enough orbital periods to inform theories of WD 1856 b's orbit. Future observations with 8-10 meter class telescopes over the next decade should achieve the sensitivity needed to detect long-period planets that might have perturbed WD 1856 b into its current orbit. 
\end{enumerate}

While we cannot yet firmly constrain the process by which WD 1856 b came to its current orbit, our work shows there is a path forward toward addressing this question. Though our best data so far indicates that WD 1856 b may be lonely, we will have to wait for more transits to  determine if it has any friends that could explain its migration.

\section*{Acknowledgements}

\bedit{We thank the anonymous referee for a very careful and helpful report.} We thank Tom Kaye for making his observations available for our analysis. This research has made use of NASA's Astrophysics Data System, the NASA Exoplanet Archive, which is operated by the California Institute of Technology, under contract with the National Aeronautics and Space Administration under the Exoplanet Exploration Program, and the SIMBAD database, operated at CDS, Strasbourg, France. The authors acknowledge the MIT SuperCloud and Lincoln Laboratory Supercomputing Center for providing high performance computing resources that have contributed to the research results reported within this paper.

S. Xu is supported by the international Gemini Observatory, a program of NSF’s NOIRLab, which is managed by the Association of Universities for Research in Astronomy (AURA) under a cooperative agreement with the National Science Foundation, on behalf of the Gemini partnership of Argentina, Brazil, Canada, Chile, the Republic of Korea, and the United States of America. 

S. Kubiak would like to thank her first computer, Freddie 1, for his service and ultimate death in trying to finish this research. His memory lives on in her new computer, Freddie 2. She would also like to thank her dog, Chevi, for the joy she brought when coding became frustrating. 

\section*{Data Availability}

\bedit{Light curves from HAO are available upon request to B. Gary \footnote{\url{bgary1@cis-broadband.com}}. Fits files containing the likelihood maps are available as supplementary materials on Zenodo\footnote{\url{https://zenodo.org/record/7682979\#.ZAoxa-zMK3K}} \citep{kubiak}. We have created a jupyter notebook demonstrating how to use these files and how to re-create the plots that we used in this paper.\footnote{\url{https://github.com/SarahKubiak/WD-1856-TTVs-Kubiak-et-al.-2023/blob/main/ReproducingPlots.ipynb}} }




\bibliographystyle{mnras}
\bibliography{refs} 








\bsp	
\label{lastpage}
\end{document}